# All Epitaxial Fabrication of a Nanowire Plasmon Laser Structure

Michael A. Mastro, Jaime A. Freitas, Jennifer K. Hite, and Charles R. Eddy, Jr.
US Naval Research Laboratory

**Abstract**
An all-epitaxial approach was demonstrated to create coaxial plasmon laser structures composed of an alumi-num plasmonic metal / SiN$_x$ dielectric / InGaN quantum well shell surrounding a p-GaN nanowire core. Strong UV lumi-nescence was observed from as-grown vertically-aligned arrays as well as horizontally-aligned nanowires transferred to a transparent carrier wafer.

**Introduction**
The plasmon laser [1] is potentially a scalable, low-threshold, efficient, sub-wavelength-size source of intense, coherent, radiation; however, achieving these characteristics is difficult owing to the inherent losses in the plasmonic metal. The key ingredient for a plasmon laser is a plasmonic metal situated at or near an interface with a semiconductor where the resonant oscillation of electrons forms a surface plasmon polariton (SPP). The SPP presents concentrated electromagnetic fields that extend into the near surface of the metal and, similarly, a few tens of nanometers into the adjacent semiconductor. In the plasmon laser, the energy injected into the adjacent semiconductor gain region is transferred to the highly localized field of the SPP. The more confined the SPP mode is within the metal, the greater the attenuation and, thus, the more gain required to overcome this loss. Incorporating feedback in addition to the gain from the semiconductor layers can then lead to SPP lasers or plasmon lasers.

Placing the semiconductor gain medium immediately adjacent to the metal, to form a single interface SPP (SISPP), will lead to significant loss via free-electron scattering in the metal as well as metal interband transitions at high energy. The loss and, consequently, necessary gain to achieve lasing can be significantly decreased by inserting a thin dielectric (i.e., insulator) between the metal and the semiconductor (MIS). Permittivity contrast at the low/high-dielectric interfaces allows strong optical confinement in the thin dielectric. This forms a tightly confined hybrid-plasmon-polariton (HPP) mode with low propagation loss. Oulton [2] demonstrated a plasmon laser formed from a CdS nanowire on a thin MgF$_2$ film on a silver substrate with strong confinement of the hybrid mode (down to $\lambda^2/400$). This hybrid mode does not have limiting cutoff and, in practice, the CdS nanowire on the MgF$_2$/silver thin-film/substrate lased at NW diameters much smaller (down to 52nm) than the minimum capable of containing a CdS photonic lasing mode (140nm). The index contrast at the ends of the structure created a Fabry-Perot cavity, with the corresponding Fabry-Perot mode spacing being observed in the emission.

A similar study [3] surrounded the perimeter of a CdS nanowire with an evaporated Ag/MgF$_2$ shell. This slight change in structure created whispering gallery modes rather than the Fabry-Perot modes of [2]. Lasing was not observed in this structure, which was attributed to losses in the metal. Nevertheless, the plasmonically enhanced nanowire provided a unique system to study hot excitons and their associated release of excess energy by emitting LO phonons, as well as possible interactions with the acoustic phonons. Above bandgap stimulation of a (plasmonic metal-free) CdS nanowire creates excitons that thermalize at a moderate rate to the k~0 momentum state from which they emit the typical free A- and B- excitons. In the plasmonic nanowire structure, the concentrated electromagnetic fields shorten the excited energy-state's relaxation rate. The addition of the plasmonic metal changed the dominant luminescence to hot (*i.e.*, non-thermalized) excitons, with the peak separation determined by the CdS LO phonon energy of 38meV.

Interfacial and surface quality is an important metric in every plasmonic structure design, [4-6] but is particularly challenging for high aspect nanowires as was observed by Oulton. [2] This paper presents a unique approach by employing an *all-epitaxial method* in which each component of the coaxial structure, *i.e.*, the InGaN-based quantum well (QW) at the pn junction, the AlN or $SiN_x$ dielectric, and the aluminum plasmonic metal, is grown in a continuous process.

**Experimental**

In this study, III-nitride nanowires were grown via a nickel nitrate seed on r-plane sapphire wafers. A 0.005 M nickel nitrate solution was repeatedly dripped onto a substrate and blown dry in $N_2$ then loaded into a modified vertical impinging flow, metal organic chemical vapor deposition reactor. A 50-Torr, $N_2/H_2$ mixed atmosphere was used during the ramp to growth temperature. Trimethylgallium was flowed for 2 sec prior to the onset of $NH_3$ flow to prevent nitridation of the nickel seeds. The initial growth in a $H_2$ ambient at a temperature of 800°C, a pressure of 50 Torr and a V/III ratio of 50 doped with $Cp_2Mg$ created the p-type nanowire core with a 100-nm cross-section. [7] The reactor temperature was lowered to 600°C and the V/III ratio was increased to 250 for the growth of a 5nm thickness $In_{0.05}Ga_{0.95}N$ shell. [8] A low percentage of InN was selected to match the hybrid-plasmon-polariton response. [9] In common III-nitride optoelectronic design, it is known that even low levels of InN incorporation can improve the radiative efficiency and improve the accumulation of holes from an adjacent p-GaN layer. [10-14]

Subsequently, the reactor temperature was increased to 675°C to grow a 5nm $SiN_x$ shell using a 0.1% $SiH_4$ source in balance hydrogen. Also demonstrated but not discussed further is that an AlN layer could be grown in place of the $SiN_x$ shell and act as an effective dielectric layer. The 5nm aluminum shell was grown at 300°C with triethylaluminum in a $H_2$ ambient. Characterization was conducted on the as-grown vertical nanowires as well as nanowires physically transferred by contact to a sapphire carrier (Figure 1).

The luminescence in nanowires was excited at room temperature with the 325-nm line of a He–Cd laser. The uncorrected photoluminescence spectra were acquired with a fiber optical spectrometer, which is fitted with an UV-extended linear CCD array and a grating blazed at 350 nm, coupled to a near-UV transmitting optical microscope. Real-color Red/Green/Blue (RGB) and single-color (R, G, or B) optical images were obtained with a near-UV CCD camera fitted with a wheel filters attached to one of the microscope ports.

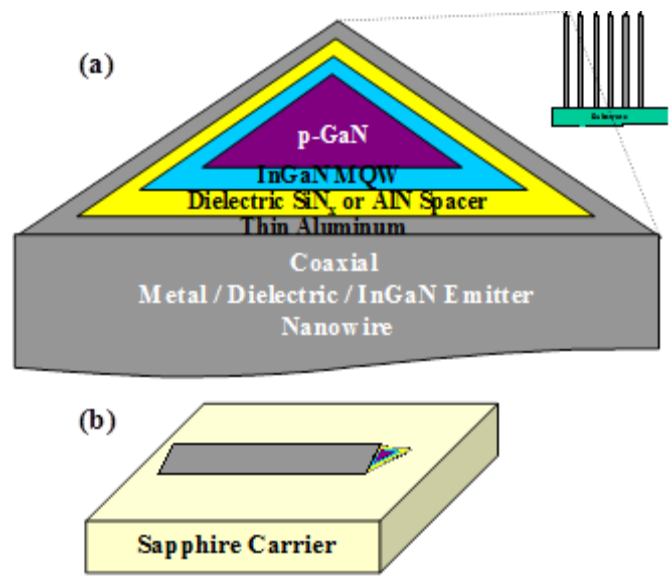

**Figure 1** (a) Vertically aligned nanowires composed of a metal / dielectric / InGaN QW shell surrounding a p-GaN core grown in an *all-epitaxial process*. (b) A physical transfer process via one-direction contact transfer was employed to horizontally align nanowires on transparent sapphire carrier substrates.

**Results**

The hybrid-plasmon-polariton was designed to coincide with the approximately 3.4eV emission of the InGaN quantum well. [9] Figure 2 displays the photo-luminescence of the as-grown vertical nanowires without and with the Al plasmonic coating. One characteristic of the plasmonic structure is to quench emission not resonant with the hybrid-plasmon-polariton. The dopant related transition near 3.1 eV is reduced for the sample with the Al thin-film. Specifically, the relative ratio of dopant to UV emission decreased by 165% after the Al coating. This change is an average over a number of nanowires in the probe beam area. Visual examination shows that the enhancement is more pronounced for nanowires oriented perpendicular to the substrate.

Figure 3 displays real color micro-photoluminescence images of as-grown nanowires with aluminum / $SiN_x$ / n-GaN / InGaN QW shells on p-GaN cores. The as-grown vertically aligned nanowires are pointing at the viewer in figure 3(a), whereas the same wires were dispersed into horizontal positions on a sapphire carrier in figure 3(b). The high density of violet dots is from the ends of nanowire structures. Yellow emission is observable in figure 3(a) from random crystallites that deposited directly on the sapphire surface. [15] Prior to the onset of growth (at elevated temperature), the nickel film manifests as a high-density of semi-spheres in a response to minimize its surface tension. It is possible for some reactants to not react with the nickel, directly reach the sapphire surface, and crystallize via unstructured nucleation.

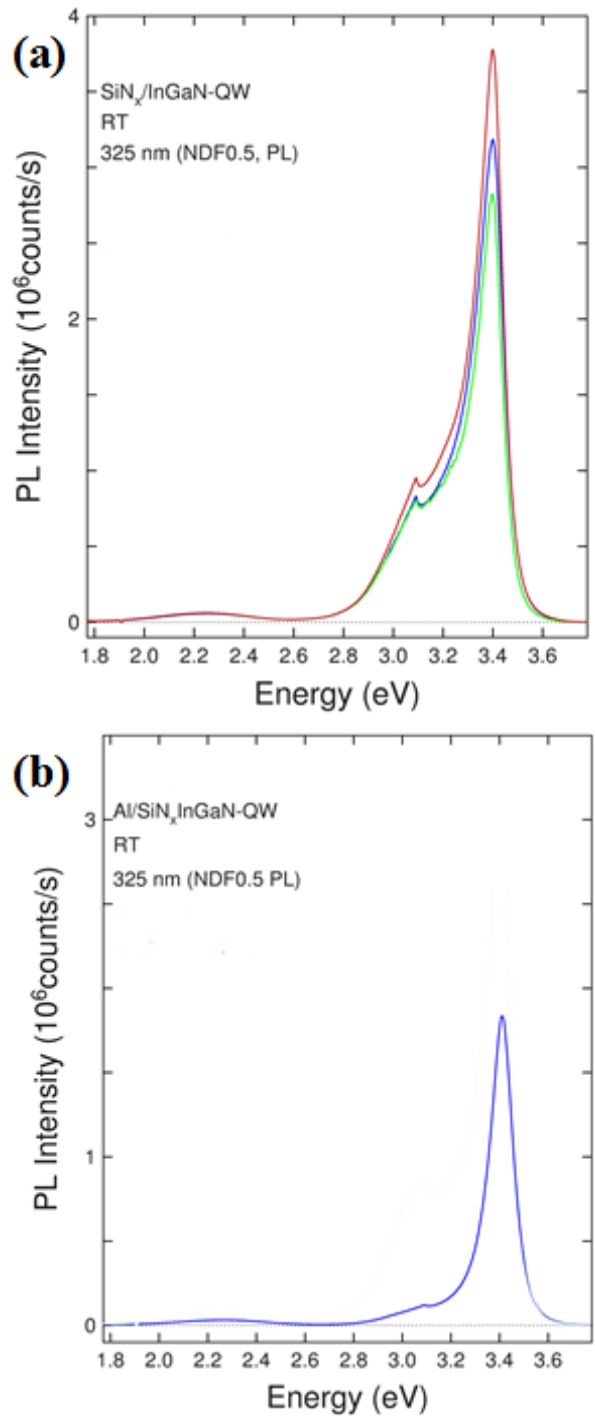

**Figure 2** Photoluminescence spectra of a $SiN_x$ / n-GaN / InGaN QW (a) without and (b) with an Al epitaxial coating. The peak response of the hybrid-plasmon-polariton coincides with the near band-edge of the InGaN well. The lower energy emission is decreased via absorption in the metal. (a) The intensity of each spectra corresponds to sample area with a proportional level of nanowire density.

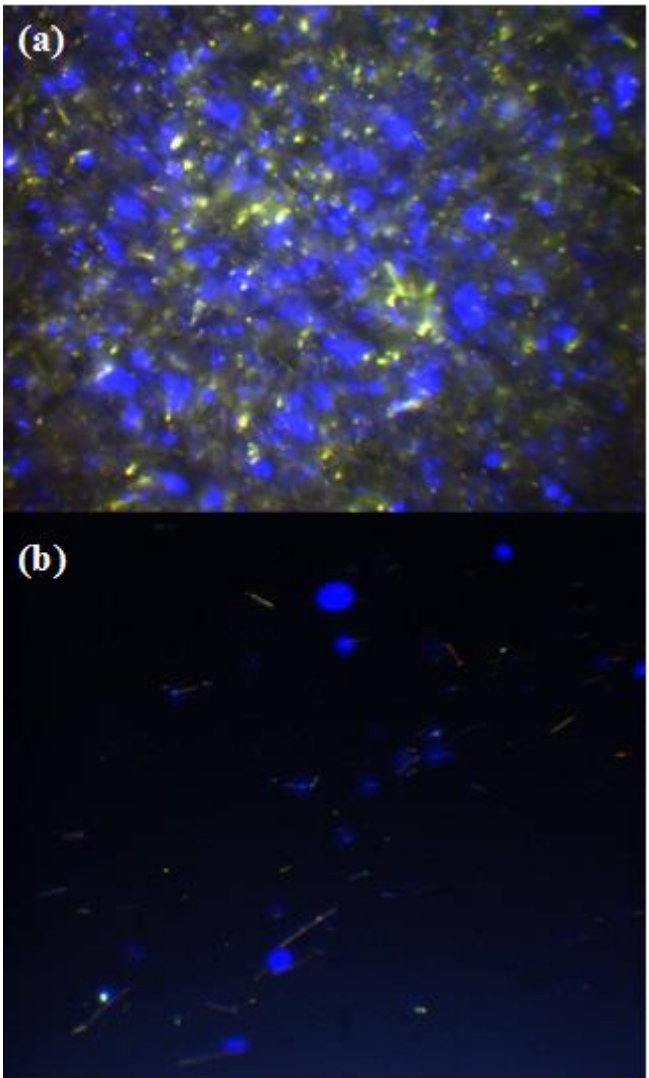

**Figure 3** Micro-photoluminescence images of (a) a high-density of vertically aligned NWs, and (b) hoziontally aligned nanowires on a sapphire carrier. The violet color emanates from the NW ends.

**Discussion**
The addition of plasmonic metals to various semiconductor systems has enhanced a number of physical effects ranging from improved absorption and emission to the alteration of subtle physical phenomena such as hot-exciton luminescence, and has enabled novel devices such as the plasmon laser. The nanowire based plasmon laser structure is promising as the vapor-liquid-solid growth mechanism is a well established technique to accurate produce multilayer nanometer-scale structures. [16] Specifically, the precise control of the layer geometries is critical to control the mode profile, and to maximize gain and minimize loss. Surface plasmons can display a large Purcell effect even in a low Q cavity. Exact location of the modal electric field maximum away from absorption layers is a key design parameter to maximize the Purcell effect. As discussed earlier, a number of plasmon laser studies have physically placed the semiconductor structure on the dielectric and/or metal film, or coated the semiconductor with low-temperature metal evaporation technique. This inherently creates an imperfect junction and thus deleterious scattering. In contrast, the all-epitaxial approach should inherently produce pristine metal / dielectric / semiconductor interfaces.

**Summary**
An *all-epitaxial* approach was used for *in-situ* deposition of the aluminum plasmonic metal / $SiN_x$ / n-GaN / InGaN QW shell on a p-GaN nanowire core. The plasmonic metal was designed to create a hybrid-plasmon-polariton that coincided with the UV emission peak of the InGaN quantum well. The ratio of resonant-UV to non-resonant emission increased from 3.75 without the Al coating to 10.2 with the plasmonic Al coating.

**Acknowledgements** Research at the Naval Research Lab is supported by the Office of Naval Research